\documentclass[12pt]{article}
 \usepackage{amssymb}
\usepackage{amsfonts}
 \usepackage{amsmath}
 \usepackage[mathscr]{eucal}
 \usepackage{amsthm}
 \usepackage{bm}
 \usepackage{graphicx}
\usepackage[T2A]{fontenc}

\numberwithin{equation}{section}
\newcommand{\be}{\begin{equation}}
\newcommand{\ee}{\end{equation}}
\newcommand{\bea}{\begin{eqnarray}}
\newcommand{\eea}{\end{eqnarray}}

\newcommand{\p}[1]{(\ref{#1})}

\newcounter{rown}

\begin{document}

\vspace*{1cm}

\begin{center}
{\LARGE\bf Massless infinite spin (super)particles and fields}

\vspace{1cm}

{\large\bf I.L. Buchbinder$^{1}$,\,\, S. Fedoruk$^2$,\,\,  A.P. Isaev$^{2,3,4}$}

\vskip 1cm

\ $^1${\it Department of Theoretical Physics,
Tomsk State Pedagogical University, \\
634041 Tomsk, Russia}, \\
{\tt joseph@tspu.edu.ru}

\vskip 0.5cm

\ $^2${\it Bogoliubov Laboratory of Theoretical Physics,
Joint Institute for Nuclear Research, \\
141980 Dubna, Moscow Region, Russia}, \\
{\tt fedoruk@theor.jinr.ru, isaevap@theor.jinr.ru}

\vskip 0.5cm

\ $^3${\it Faculty of Physics, Lomonosov Moscow State University,
119991 Moscow, Russia}

\vskip 0.5cm

\ $^4${\it St.Petersburg Department of Steklov Mathematical Institute of RAS, \\ Fontanka 27, 191023 St. Petersburg, Russia}

\end{center}

\vspace{0.5cm}

\begin{abstract}
\noindent A new twistorial field formulation of a
massless infinite spin particle is derived. We find a twistorial infinite spin field and derive
its helicity decomposition. The twistorial equations of motion for infinite spin fields in the
cases of integer and half-integer helicities are derived. We show
that the infinite integer-spin field and infinite half-integer-spin
field form the $\mathcal{N}{=}\,1$ infinite spin supermultiplet. The corresponding
supersymmetry transformations are presented.
We prove that the supersymmetry algebra is closed on-shell.
\end{abstract}

\vspace{1cm}


\phantom{Keywords: }

\vspace{1cm}

\begin{center}
{\it Contribution to the Volume dedicated to the 80-th Anniversary Jubilee \\ of Andrei A. Slavnov}
\end{center}

\setcounter{footnote}{0}
\setcounter{equation}{0}

\newpage

\section{Introduction}

Principles of symmetry, formulated in terms of group
theory, play a remarkable role in theoretical and mathematical
physics. It is sufficient to say that the Standard Model is
constructed on the base of gauge principle, which imposes the
essential restrictions both on classical Lagrangian and on
scattering amplitudes. The fundamental contribution to realization
of the gauge principle in quantum field theory was done by A.A.\,Slavnov \cite{Slavnov}. The paper under consideration is devoted to
some aspects of symmetry related to Poincar\'{e} group.

Relativistic symmetry associates the elementary particles
with irreducible representations of the Poincar\'{e} group
$ISO^{\uparrow}(1,3)$ (or its covering $ISL(2,\mathbb{C})$).
Classification of the $ISO^{\uparrow}(1,3)$ unitary irreducible
representations was given in \cite{Wigner39,Wigner47,BargWigner}.
Unitary irreducible representations of the Poincar\'{e} group, which
are usually interesting from the physical point of view,
act in the space of states with non-negative mass squared ${\sf m}^2 \geq 0$
and non-negative energy $E = k_0 \geq 0$ (here $k_0$ is zero
component of 4-momentum of particle).

To characterize these irreducible representations we need to consider
the corresponding irreducible representations of the Lie algebra  {$iso(1,3)$} with
generators $\hat{P}_n, \hat{M}^{mk} $
(components of momentum and angular momentum)
and defining relations
\begin{equation}\label{P-al}
\begin{array}{l}
[\hat{P}_n , \, \hat{P}_m] = 0 \; , \qquad
[\hat{P}_n , \, \hat{M}_{m k}] = i \,(  \eta_{kn} \hat{P}_m -
 \, \eta_{m n} \hat{P}_k )  \; , \\ [6pt]
\, [\hat{M}_{nm}, \, \hat{M}_{k\ell}] =
i \,( \eta_{nk} \hat{M}_{m\ell} - \eta_{mk} \hat{M}_{n\ell}
+ \eta_{m\ell} \hat{M}_{nk} - \eta_{n\ell} \hat{M}_{mk} ) \; ,
\end{array}
\end{equation}
where metric tensor is $||\eta_{mk}|| =  {\rm diag}(+1,-1,-1,-1) $.

There are {{\bf 2 classes}} of physically  interesting
 unitary irreducible representations (irreps) of Poincar\'{e} group:
{\bf 1.} massive irreps and  {\bf 2.} massless irreps.
\begin{description}
\item[1.\,Massive irreps.]
The algebra {{$iso(1,3)$}} has two Casimir operators
$\hat{P}^n \hat{P}_n$ and $\hat{W}^n \hat{W}_n$, where
$$
\hat{W}_n = \frac{1}{2} \, \varepsilon_{nmkr} \,
\hat{M}^{mk} \, \hat{P}^r 
$$
are components of the Pauli-Lubanski vector which satisfy
$$
\hat{W}_n \hat{P}^n = 0 \, , \qquad
[\hat{W}_k, \; \hat{P}_n] = 0 \, , \qquad
[\hat{W}_m, \, \hat{W}_n] =  i \, \varepsilon_{mnkr} \hat{W}^k \, \hat{P}^r \, .
$$
On the space of states of massive irreducible representations the Casimir operators
are proportional to the unite operator $\mathbb{I}$:
$$
\hat{P}^n \hat{P}_n = {\sf m}^2 \; \mathbb{I}
\;\;\; ({\sf m}^2>0)  , \qquad
\hat{W}^n \hat{W}_n = - {\sf m}^2 \, j(j+1)\; \mathbb{I} \; ,
$$
where the real number
  {{${\sf m} > 0$}} is called
  {{mass}} and the real number
  {{$j \in \mathbb{Z}_{\geq 0}/2$}} is called
  {{spin}}.
\item[2.\,Massless irreps.]
The Casimir operators of $iso(1,3)$ are
$$
 \hat{P}^n \hat{P}_n = {\sf m}^2 = 0 \; , \qquad
 \hat{W}^2 = \hat{W}^n \hat{W}_n = - \mu^2  \; .
$$
In this case we have two subcases: {\bf A)} {{$\mu^2=0$}}
and {\bf B)} {{$\mu^2 \neq 0$}}.
\begin{description}
\item[In massless case A)]
we obtain usual massless helicity representations with
$$
\hat{W}^2  = 0, \quad \hat{P}^2 = 0, \quad
\hat{P}_n \hat{W}^n  = 0
 \qquad \stackrel{{\rm for} \;\; \mathbb{R}^{1,3}}{\Longrightarrow}
  \qquad \hat{W}_n =
 \hat{\Lambda} \cdot \hat{P}_n \; ,
$$
where central element $\hat{\Lambda} \in iso(1,3)$ is called
helicity  operator and its eigenvalues are  $\Lambda = 0,\pm 1/2,\pm
1, \pm 3/2, \dots$.

\item[In massless case B)] we have
$$
   \hat{W}^2  = -\mu^2, \quad \hat{P}^2 = 0, \quad
 \hat{P}_n \hat{W}^n  = 0 \; ,
  $$
which correspond to massless irreducible representation
 of infinite (continuous) spin.
To describe these representations we have to introduce ``canonically
conjugate'' to $\hat{P}_k$, $\hat{W}_n$ variables $x_k$,
$y_n$:
$$
  x = (x_0,x_1,x_2,x_3) \in \mathbb{R}^{1,3} \, , \qquad
 y = (y_0,y_1,y_2,y_3) \in \mathbb{R}^{1,3} \, .
$$
Then, as it was shown in  \cite{Wigner39,Wigner47,BargWigner}, the
massless infinite spin irreducible representations of the
Poincar\'{e} group are realized in the space of wave
functions (WF) $\Phi(x,y)$ which satisfy the conditions
\begin{equation} \label{one}
\begin{array}{l}
{\displaystyle \frac{\partial}{\partial x^m}\frac{\partial}{\partial x_m}
\ \Phi = 0 \,, \qquad \frac{\partial}{\partial x^m}\frac{\partial}{\partial y_m} \
\Phi = 0 \,,   } \\ [7pt]
{\displaystyle \frac{\partial}{\partial y^m}\frac{\partial}{\partial y_m}
\ \Phi = \mu^2\,  \Phi\,,  \qquad -\,i\,y^m \,\frac{\partial}{\partial x^m} \ \Phi = \Phi } \, .
\end{array}
\end{equation}
\end{description}
\end{description}

This paper is devoted to some aspects of theory of {{massless infinite (or continues) spin
unitary irreducible representations of the $ISL(2,\mathbb{C})$}} group.
Various problems related to the quantum-mechanical and field descriptions of
such states were considered in a wide range of works devoted to particles and fields of infinite (continues) spin (see, e.g.,
\cite{Iv-Mack}--\cite{Metsaev19}). Motivation of the investigations of the infinite spin representations is
caused by an identical spectrum of states of the infinite spin
theory \cite{Iv-Mack} and the higher-spin theory
\cite{Vas1989,Vas1991,Vas1992} (see also the reviews
\cite{Vas2001,Vas2002,Vas2005}) and by its potential relation to
the string theory (see \cite{MundSY} and recent paper \cite{Vas2018} and
references in it) as candidates for Quantum Gravity Theory.

In our recent papers \cite{BFIR,BFI} we constructed a new model of an infinite (continuous)
spin particle, which is a generalization of  the twistor formulation of
standard (with fixed helicity) massless particle
\cite{Pen67,PC72,PenRin} to massless infinite spin representations.
As a result of a quantization procedure,
 we obtained infinite spin fields that demonstrate a
transparent decomposition of continues spin irreducible
 representations into infinite sum of states with all helicities.
  We stress that for massless case {\bf B}, irreducible
 representations are not characterized by definite helicities.
Besides, using
the field twistor transform, we can now get the
space-time--spinorial description of infinite spin fields with
integer or half-integer helicities that form
 a supermultiplet of infinite spins \cite{BKRX,Zin}.

  This paper is based on the results obtained in \cite{BFIR,BFI}.

\section{Wigner-Bargmann space-time formulation}

The Wigner-Bargmann space-time formulation
\cite{Wigner39,Wigner47,BargWigner} of the irreducible infinite spin
massless representation can be realized by means of
quantization of the particle model with the following Lagrangian
\begin{equation}\label{L-sp-t}
 {\cal L}_{sp.-time}\ =\ p_m \dot x^m \ +\ q_m \dot y^m \ +\ e \, p_m p^m
\ +\ e_1 \, p_m q^m \ +\ e_2 \, \left(q_m q^m +\mu^2\right) \ +\
 e_3 \, \left(p_m y^m -1\right) ,
\end{equation}
where $\{ p_n(\tau), \; q_n(\tau) \}$
are momenta canonically conjugated to
 coordinates {{$\{ x_n(\tau), \; y_n(\tau) \}$}},
$\tau$ is the evolution parameter and
$\dot{x}_k(\tau):= \partial_{\tau}x_k(\tau)$.
The Lagrangian \p{L-sp-t} yields the canonical Poisson brackets
$$
 \left\{ x^m, p_n \right\}=\delta^m_n\,,\qquad
\left\{y^m, q_n \right\}=\delta^m_n
$$
and first-class constraints
\begin{equation}\label{con-sp-t}
\begin{array}{c}
T := p_m p^m \ \approx \ 0 \,, \qquad
T_1 := p_m q^m \ \approx \ 0  \,,  \\ [6pt]
T_2 := q_m q^m +\mu^2 \ \approx \ 0  \,,  \qquad
T_3 := p_m y^m -1 \ \approx \ 0 \,,
\end{array}
\end{equation}
which correspond to the Wigner-Bargmann equations (\ref{one}).
The variables $e(\tau)$, $e_1(\tau)$, $e_2(\tau)$, $e_3(\tau) $ are Lagrange multipliers for the constraints \p{con-sp-t}.
Nonvanishing Poisson brackets of the constraints \p{con-sp-t} are
$$
\left\{ T_1, T_3 \right\}=-T\,,\qquad \left\{T_2, T_3\right\}=-2T_1\,.
$$

The action $\displaystyle S_{sp.-time}=\int d\tau {\cal L}_{sp.-time}$
is invariant under the transformations which generated by the quantities
$$
P_m = p_m \,, \qquad
M_{mn} = x_m p_n - x_n p_m + y_m q_n - y_n q_m \, .
$$
These charges form the Poincar\'{e} algebra with respect to Poisson brackets.
 We see that additional coordinates
 $y^m$ in the arguments of these fields play the role of
spin variables.

Now by making use of constraints
$T\approx0$, $T_1\approx0$, $T_2\approx0$, $T_3\approx0$ we obtain relations
$$
P_m P^m \approx 0 \, , \qquad
 W_m W^m=\frac12\, M_{nk}M^{nk}P_m P^m-
  M_{mk}M^{nl}P^k P_l\approx -\mu^2\,.
$$
where $W_m=\frac12\,\varepsilon_{mnkl}P^n M^{kl}$ are
the components of the Pauli-Lubanski pseudovector.
Therefore, the model with Lagrangian
${\cal L}_{sp.-time}$ indeed
 describes the massless particle with continuous spin. We note that vectors
$q_m$ and  $W_m=\varepsilon_{mnkl}p^n y^{l}q^{l}$ do not coincide to each other
 and components $W_m$ strictly speaking
 are not canonically conjugated to $y_m$.

After canonical quantization the constraints
\p{con-sp-t}  yield the Wigner-Bargmann equations \p{one} for the continuous spin fields
{{$\Phi(x,y)$}}.

\section{\bf Twistorial formulation of continuous spin particles.}

Our aim is to reformulate the Wigner-Bargmann model with Lagrangian
(\ref{L-sp-t}) in terms of twistor variables. More precisely we need to construct a twistor particle model which is
classically equivalent to the Wigner-Bargmann model with Lagrangian
${\cal L}_{sp.-time}$.

Below we will use the following two-spinor  conventions about
notation. The totally antisymmetric tensor $\epsilon^{mnkl}$ has the
component {{$\epsilon^{0123}=1$}}.
 We use the set of $\sigma$-matrices: $\sigma^n =
(\sigma^0\equiv I_2,\, \sigma^1,\, \sigma^2,\, \sigma^3)$
 and the set of dual
$\sigma$-matrices: $\tilde{\sigma}^n =(\sigma^0,-\sigma^1,
 -\sigma^2,-\sigma^3)$, where  {{$\sigma^i$}}
 are usual Pauli matrices.
 We also use standard
van der Waerden spinor notation with dotted and undotted spinor
indices and raise and lower them by means of metrics:
$\epsilon_{\alpha\beta}$, $\epsilon_{\dot\alpha\dot\beta}$ and their
inverse $\epsilon^{\alpha\beta}$, $\epsilon^{\dot\alpha\dot\beta}$
with components $\epsilon_{12}=-\epsilon_{21}=1$. In particular
$(\tilde\sigma_{m})^{\dot\alpha\beta}=
\epsilon^{\dot\alpha\dot\delta}\epsilon^{\beta\gamma}
(\sigma_m)_{\gamma\dot\delta}$. The links between the Minkowski
four-vectors and spinorial quantities are
$$
A_{\alpha\dot\beta}={\textstyle\frac{1}{\sqrt{2}}}\,
A_m(\sigma^m)_{\alpha\dot\beta} \, , \qquad
A^{\dot\alpha\beta}={\textstyle\frac{1}{\sqrt{2}}}\,
A_m(\tilde\sigma^m)^{\dot\alpha\beta} \, , \;\;
A_m
={\textstyle\frac{1}{\sqrt{2}}}\,A_{\alpha\dot\beta}
(\tilde\sigma_m)^{\dot\beta\alpha}\,,
$$
so that $ A^m B_m=A_{\alpha\dot\beta}B^{\dot\beta\alpha}$.

In \cite{BFIR,BFI} we propose that
the twistorial formulation of the infinite (continuous) spin particle is described by
the bosonic Weyl spinors
\begin{equation}\label{tw-sp-1}
\pi_{\alpha} \; , \;\;\; \bar\pi_{\dot\alpha}:=(\pi_{\alpha})^* \; , \qquad
\rho_{\alpha} \; , \;\;\;
\bar\rho_{\dot\alpha}:=(\rho_{\alpha})^*\; ,
\end{equation}
and their canonically conjugated spinors
\begin{equation}\label{tw-sp-2}
 \omega^{\alpha}\; , \;\;\; \bar \omega^{\dot\alpha}:=(\omega^{\alpha})^* \; , \qquad\eta^{\alpha} \; , \;\;\; \bar{\eta}^{\dot\alpha}:=(\eta^{\alpha})^* \; .
\end{equation}
The nonzero Poisson brackets of these spinors are
$$
\left\{ \omega^{\alpha}, \pi_{\beta} \right\}=
\left\{ \eta^{\alpha}, \rho_{\beta} \right\}=\delta^\alpha_\beta\,,\qquad
\left\{\bar \omega^{\dot\alpha}, \bar\pi_{\dot\beta} \right\}=\left\{\bar \eta^{\dot\alpha}, \bar\rho_{\dot\beta} \right\}=\delta^{\dot\alpha}_{\dot\beta}\,.
$$
Twistorial Lagrangian of the infinite (continuous) spin particle is
written in the form \cite{BFIR,BFI}:
\begin{equation}
\label{two}
{\cal L}_{twistor}= \pi_{\alpha}\dot \omega^{\alpha} \
+ \ \bar\pi_{\dot\alpha}\dot{\bar \omega}^{\dot\alpha}  \ + \
\rho_{\alpha}\dot \eta^{\alpha}  \ + \
\bar\rho_{\dot\alpha}\dot{\bar \eta}^{\dot\alpha} \ + \
l\,{\mathcal{M}} \ + \  k\,{\mathcal{U}} \ + \  \ell\,{\mathcal{F}}
\ +\  \bar\ell\,\bar{\mathcal{F}}\,,
\end{equation}
where $l(\tau)$, $k(\tau)$, $\ell(\tau)$, $\bar\ell(\tau)$
are the Lagrange multipliers for the constraints
\begin{eqnarray}
\mathcal{M} &:=& \pi^{\alpha}\rho_{\alpha}\,\bar\rho_{\dot\alpha}\bar\pi^{\dot\alpha}
-\mu^2/2 \ \approx\  0\,,
\label{tw-const-1} \\ [6pt]
{\mathcal{F}}&:=& \eta^{\alpha}\pi_{\alpha}-1\approx 0\,,\qquad\quad
\bar{\mathcal{F}}\ :=\ \bar\pi_{\dot\alpha}\bar \eta^{\dot\alpha}-1\approx 0\,,
\label{tw-const-2} \\  [6pt]
{\mathcal{U}} &:=& i \; (\omega^{\alpha}\pi_{\alpha}-\bar\pi_{\dot\alpha}\bar \omega^{\dot\alpha}
+\eta^{\alpha}\rho_{\alpha}-\bar\rho_{\dot\alpha}\bar \eta^{\dot\alpha})
\approx 0\,, \label{tw-const-3}
\end{eqnarray}

 One can check that the first-class constraints \p{tw-const-1}, \p{tw-const-1}, \p{tw-const-1} generate
 abelian Lie group which acts in the phase space of spinors
  (\ref{tw-sp-1}), (\ref{tw-sp-2}) as follows:
 \begin{equation}
 \label{cantr1}
  \left(\!\!
 \begin{array}{cc}
 \pi_1 \! & \! \rho_1 \\
 \pi_2 \! & \! \rho_2
 \end{array}
 \!\!\right) \rightarrow
 \left(\!\!
 \begin{array}{cc}
 \pi_1 \! & \! \rho_1 \\
 \pi_2 \! & \! \rho_2
 \end{array}
\! \!\right)  \left(\!\!
 \begin{array}{cc}
 e^{i\beta} \! & \! \alpha e^{i\beta}\\
 0 \! & \! e^{i\beta}
 \end{array}
 \!\!\right) \, ,
 \end{equation}
 \begin{equation}
 \label{cantr2}
 \left(\!
 \begin{array}{cc}
 \eta_1 \! & \! \omega_1 \\
 \eta_2 \! & \! \omega_2
 \end{array}
 \!\!\right)
  \rightarrow  \left(\!\!
 \begin{array}{cc}
 \eta_1 \! & \! \omega_1 \\
 \eta_2 \! & \! \omega_2
 \end{array}
 \!\!\right)  \left(\!\!
 \begin{array}{cc}
 e^{-i\beta} \! & \! -\alpha e^{-i\beta}\\
 0 \! & \! e^{-i\beta}
 \end{array}
 \!\!\right) +  \frac{2}{\mu^2}
 (\bar\rho_{\dot\alpha}\bar\pi^{\dot\alpha})  \left(\!\!
 \begin{array}{cc}
 \pi_1 \! & \! \rho_1 \\
 \pi_2 \! & \! \rho_2
 \end{array}
 \!\!\right) \left(\!\!
 \begin{array}{cc}
 \gamma  \! & \! 0 \\
 0 \! & \! -\gamma
 \end{array}
 \!\!\right)
 \end{equation}
and the transformations that are complex conjugation of \p{cantr1}, \p{cantr2},
where $\beta(\tau) , \gamma(\tau) \in \mathbb{R}$
 and
$\alpha(\tau) \in \mathbb{C}\backslash 0$
 are the parameters of the gauge group generated
 by constraints \p{tw-const-1}, \p{tw-const-1}, \p{tw-const-1}.

{\bf Proposition 1.}
{\em The Wigner-Bargmann space-time (\ref{L-sp-t}) and twistorial (\ref{two})
formulations of the infinite (continuous) spin particle are equivalent on the
classical level by means of the generalized Cartan-Penrose relations \cite{Pen67,PC72,PenRin}
\begin{equation}\label{tw-CP-1}
p_{\alpha\dot\beta}=\pi_{\alpha}\bar\pi_{\dot\beta}\,,
\qquad
q_{\alpha\dot\beta}=\pi_{\alpha}\bar\rho_{\dot\beta}
+\rho_{\alpha}\bar\pi_{\dot\beta}\,,
\end{equation}
and by the following generalized incidence relations \cite{Pen67,PC72,PenRin}:
}
\begin{equation}\label{tw-inc-1}
\omega^{\alpha}=\bar\pi_{\dot\alpha}x^{\dot\alpha\alpha}
+\bar\rho_{\dot\alpha}y^{\dot\alpha\alpha}\,,\qquad
\bar \omega^{\dot\alpha}=x^{\dot\alpha\alpha}\pi_{\alpha}
+y^{\dot\alpha\alpha}\rho_{\alpha}\,,
\end{equation}
\begin{equation}\label{tw-inc-2}
\eta^{\alpha}=\bar\pi_{\dot\alpha}y^{\dot\alpha\alpha}\,,\qquad
\bar \eta^{\dot\alpha}=y^{\dot\alpha\alpha}\pi_{\alpha}\,.
\end{equation}
The proof of this Proposition is straightforward
 and is given in  \cite{BFIR,BFI}.

\section{Quantization of the twistorial model and
twistor field of the infinite spin particle}

Quantization of the model is drastically simplified if we introduce new spinorial
 variables by means of Bogolyubov canonical transformations
 (cf. gauge transformations (\ref{cantr1}),
  (\ref{cantr2})):
$$
  \begin{array}{c}
 \left(\!\!
 \begin{array}{cc}
 \pi_1 \! & \! \rho_1 \\
 \pi_2 \! & \! \rho_2
 \end{array}
 \!\!\right) 
 = \sqrt{M} \left(\!\!
 \begin{array}{cc}
 p^{(z)}_1 \! & \! 0 \\
p^{(z)}_2 \! &  p^{(s)}/p^{(z)}_1
 \end{array}
\! \!\right)  \left(\!\!
 \begin{array}{cc}
 1 \! & \! p^{(t)} \\
 0 \! & \! 1
 \end{array}
 \!\!\right) \, , \;\;
   \end{array}
 $$
 $$
 \left(\!\!
 \begin{array}{cc}
 \eta_1 \! & \! \omega_1 \\
 \eta_2 \! & \! \omega_2
 \end{array}
 \!\!\right) 
 = \left(\!\!
 \begin{array}{cc}
 0 \! & \!  z_1/\sqrt{M} \\
 -t/\pi_1 \! &   z_2/\sqrt{M}
 \end{array}
\! \!\right)  \left(\!\!
 \begin{array}{cc}
 1 \! & \! - p^{(t)} \\
 0 \! & \! 1
 \end{array}
 \!\!\right) +
 \frac{s}{M}  \left(\!\!
 \begin{array}{cc}
 \pi_1 \! & \! \rho_1 \\
 \pi_2 \! & \! \rho_2
 \end{array}
 \!\!\right)   \left(\!\!
 \begin{array}{cc}
 1 \! & \! 0 \\
 0 \! & \! -1
 \end{array}
 \!\!\right)\, , \;\;
 $$
 where $M = \mu/\sqrt{2}$ and new variables are defined by the expressions
\begin{equation}\label{n-var}
\begin{array}{c}
{\displaystyle p^{(z)}_{\alpha} = \pi_{\alpha}/\sqrt{M}\,,\qquad
p^{(s)}=\pi^{\alpha}\rho_{\alpha}/M\,,\qquad
p^{(t)}=\rho_{1}/\pi_{1} \; , }\\ [6pt]
{\displaystyle \omega^\alpha =  \,\frac{1}{\sqrt{M}}\,z^{\alpha} \ - \
\frac{1}{M}\,s\,\rho^{\alpha}  \ - \
\frac{\delta^{\alpha 1}}{\pi_1}\, t\, p^{(t)}  \,,}\\[6pt]
{\displaystyle \eta^\alpha =  \ \,
\frac{1}{M}\,s\,\pi^{\alpha}  \ + \ \frac{\delta^{\alpha 1}}{\pi_1}\, t  \, .}
\end{array}
\end{equation}
By means of
complex conjugation we obtain the conjugated coordinates $\bar{z}^{\dot\alpha}$, $\bar{s}$, $\bar{t}$ and their momenta $\bar{p}^{(z)}_{\dot\alpha}$, $\bar{p}^{(s)}$, $\bar{p}^{(t)}$.
The nonzero canonical Poisson brackets of the new variables are
$$
\left\{ z^{\alpha}, p^{(z)}_{\beta} \right\}= \delta^\alpha_\beta\,, \quad
\left\{\bar z^{\dot\alpha}, \bar p^{(z)}_{\dot\beta} \right\}=
\delta^{\dot\alpha}_{\dot\beta}\,,\qquad
\left\{ s, p^{(s)} \right\}= \left\{\bar s, \bar p^{(s)} \right\}=1\,,\qquad
\left\{ t, p^{(t)} \right\}= \left\{\bar t, \bar p^{(t)} \right\}=1\, .
$$
In terms of new variables \p{n-var} the constraints \p{tw-const-1}, \p{tw-const-2}, \p{tw-const-3} of spinorial model \p{two}
take very simple form
\begin{eqnarray}
{\mathcal{M}}^\prime &:=&  p^{(s)} \bar{p}^{(s)} -1\ \approx\  0\,,
\label{tw-const-1a}\\ [6pt]
{\mathcal{F}}^\prime &:=&  t - 1\ \approx\ 0\,,\qquad \bar{\mathcal{F}}^\prime \ := \ \bar{t} - 1\ \approx\  0\,,
\label{tw-const-2a}\\ [6pt]
{\mathcal{U}}^\prime &:=& \frac{i}{2}\,\left(z^{\alpha}p^{(z)}_{\alpha} - \bar z^{\dot\alpha}\bar p^{(z)}_{\dot\alpha}\right) \ +
\ i \left( s p^{(s)} - \bar s\bar p^{(s)} \right)
\ \approx\ 0\,.
\label{tw-const-3a}
\end{eqnarray}
After canonical quantization
 {{$[.,.] = i \, \{.,.\} $}}
these constraints turn into equations of motion
\begin{eqnarray}
&& \left(p^{(s)} \bar{p}^{(s)} -1\right) \Psi^{(c)} \ =\  0\,, \label{tw-eq-1}
\\ [5pt]
&&\frac{\partial}{\partial p^{(t)}}\ \Psi^{(c)}\ =\
\frac{\partial}{\partial \bar{p}^{(t)}} \ \Psi^{(c)}\ =\ -i\,\Psi^{(c)}\,,
\label{tw-eq-2} \\ [5pt]
&& \left[\frac{1}{2}\,\left(p^{(z)}_{\alpha}\frac{\partial}{\partial p^{(z)}_{\alpha}} \ -\
\bar{p}^{(z)}_{\dot\alpha}\frac{\partial}{\partial \bar{p}^{(z)}_{\dot \alpha}}\right) \ +\
p^{(s)} \frac{\partial}{\partial p^{(s)}} \ - \
\bar{p}^{(s)}\frac{\partial}{\partial \bar{p}^{(s)} }
\right] \Psi^{(c)}\ =\  c\ \Psi^{(c)}\,, \label{tw-eq-3}
\end{eqnarray}
 where differential operators in their left hand sides
 are  quantum counterparts of the constraints  \p{tw-const-1a}, \p{tw-const-2a}, \p{tw-const-3a}. In equations \p{tw-eq-1}, \p{tw-eq-2}, \p{tw-eq-3}
 wave function (or
spinorial field)
$$
\Psi^{(c)}\big( p^{(z)}_{\alpha},\, \bar{p}^{(z)}_{\dot\alpha}; \,
p^{(s)}, \, \bar{p}^{(s)}; \, p^{(t)}, \, \bar{p}^{(t)}\big) \; ,
$$
 is taken in ''momentum representation'' and
describes physical states, which form the
space of irreducible representation of Poincar\'{e} group with continues spin.
The constant
$c$ is related to the ambiguity of operator ordering
in equation \p{tw-eq-3}. In other words,
constant $c$ is an analog of the vacuum energy in the quantum
oscillator model.

Equations of motion \p{tw-eq-1}, \p{tw-eq-2} can be solved
explicitly in the form
\begin{equation}\label{wf-red}
\Psi^{(c)} \ = \  \delta\left(p^{(s)}\cdot \bar{p}^{(s)}-1\right)\; e^{-i(p^{(t)}+\bar{p}^{(t)})}
\,\sum\limits_{k=-\infty}^{\infty} e^{-ik \varphi} \,
\tilde\psi^{(c+k)}\big( p^{(z)},\, \bar{p}^{(z)}\big) \,,
\end{equation}
where {{$e^{i\varphi} := (p^{(s)}/\bar{p}^{(s)})^{1/2}$}}.
Due to the constraint \p{tw-eq-3} the coefficient functions
 $\tilde\psi^{(c+k)}( p_z,\, \bar{p}_z)$ satisfy the equations
\begin{equation}\label{wf-eq-hom}
\frac{1}{2}\,\left(p^{(z)}_{\alpha}\frac{\partial}{\partial p^{(z)}_{\alpha}} \ -\
\bar{p}^{(z)}_{\dot\alpha}\frac{\partial}{\partial \bar{p}^{(z)}_{\dot \alpha}}\right)
\tilde\psi^{(c+k)}
 \ =\  \big(c+k\big) \, \tilde\psi^{(c+k)} \,.
\end{equation}

Now we can restore the dependence of the wave function \p{wf-red}
on the twistor variables. As result we obtain the following statement. \\
{\bf Proposition 2.}
{\em The twistor wave function which is general solution
 of the equations of motion \p{tw-eq-1}, \p{tw-eq-2}, \p{tw-eq-3}
 is represented in the form
\begin{equation}\label{tw-wf-gen}
\Psi^{(c)}( \pi,\bar \pi;\rho,\bar\rho) \ = \
\delta\left((\pi\rho)(\bar\rho\bar\pi)-M^2\right)\,
e^{\displaystyle -i\left(\frac{\rho_{1}}{\pi_{1}}+
\frac{\bar\rho_{1}}{\bar\pi_{1}}\right)}\,
\hat{\Psi}^{(c)}( \pi,\bar \pi;\rho,\bar\rho)\,,
\end{equation}
where we make use the shorthand notation {{$(\pi\rho):=\pi^{\beta}\rho_{\beta}$,
$(\bar\rho\bar\pi):=\bar\rho_{\dot\beta}\bar\pi^{\dot\beta}$}} and
\begin{equation}\label{tw-wf-gen1}
\hat{\Psi}^{(c)}( \pi,\bar \pi;\rho,\bar\rho)  =  \psi^{(c)}( \pi,\bar \pi) +
\sum\limits_{k=1}^{\infty} (\bar\rho\bar\pi)^{k}\,
\psi^{(c+k)}
+ \sum\limits_{k=1}^{\infty} (\pi\rho)^{k}\,
\psi^{(c-k)}\, .
\end{equation}
The coefficient functions
{{$\psi^{(c\pm k)}\big( \pi,\bar \pi\big)$}}
obey the condition
\begin{equation}\label{tw-wf-gen1-hel}
\Lambda \cdot
\psi^{(c\pm k)}( \pi,\bar \pi)\ =\  - \big(c\pm k\big)\,
\psi^{(c\pm k)}( \pi,\bar \pi)\, ,
\end{equation}
where $\displaystyle \Lambda = - \frac{1}{2}\,\left(\pi_{\alpha}\frac{\partial}{\partial \pi_{\alpha}} \ -\
\bar \pi_{\dot\alpha}\frac{\partial}{\partial \bar
\pi_{\dot \alpha}}\right)$ is the helicity operator.
}

In view of condition (\ref{tw-wf-gen1-hel}),
 to describe the bosonic infinite spin representation
 related to all integer helicities, we put
$$
c=0
$$
and therefore consider the twistorial field
$\Psi^{(0)} ( \pi,\bar \pi;\rho,\bar\rho)$. Note that
 complex conjugate field $\bar\Psi^{(0)}$ also has zero charge $c=0$.
Similarly, to describe the infinite spin representation related to
half-integer helicities we take for $c$ the value
$$
c=- \,\frac12 \, .
$$
In view of condition (\ref{tw-wf-gen1-hel}),
the corresponding wave function $\Psi^{(-1/2)} ( \pi,\bar \pi;\rho,\bar\rho)$
contains in its expansion only half-integer helicities.
The complex conjugate field
$\bar\Psi^{(+1/2)} ( \pi,\bar \pi;\rho,\bar\rho)$
possesses the charge $c=+1/2$.

\vspace{0.2cm}

\noindent
{\bf Proposition 3.} {\em The twistor wave function
$\Psi^{(c)}( \pi,\bar \pi;\rho,\bar\rho)$,
defined in {\bf \em Proposition 2}, describes the massless particle of the infinite (continuous) spin:
 \begin{equation}\label{WW1}
W^{\alpha\dot\gamma}W_{\alpha\dot\gamma}\cdot \Psi^{(c)} \ = \
 -\mu^2\,\Psi^{(c)} \,,
 \end{equation}
where $W_{\alpha\dot\gamma} =\frac{1}{\sqrt{2}} W_m (\sigma^m)_{\alpha\dot\gamma}$ is the Pauli-Luba\'{n}ski operator
\begin{equation}\label{WW2}
W_{\alpha\dot\gamma}=
\pi_{\alpha}\bar\pi_{\dot\gamma}\, \Lambda
+\frac{1}{2}\left[\pi_{\alpha}\bar\rho_{\dot\gamma} \Big(\bar\pi_{\dot\beta}\,\frac{\partial}{\partial
\bar\rho_{\dot\beta}}\Big)
-\rho_{\alpha}\bar\pi_{\dot\gamma} \Big(\pi_{\beta}\,\frac{\partial}{\partial
\rho_{\beta}}\Big)\right]
+\frac{1}{2}\left[(\bar\rho\bar\pi)\,\pi_{\alpha}\,
\frac{\partial}{\partial\bar\rho^{\dot\gamma}}- (\pi\rho)\,\bar\pi_{\dot\gamma}\,\frac{\partial}{\partial
\rho^{\alpha}}\right]\,.
\end{equation}}
{\bf Proof.}  Substitute (\ref{WW2}) into (\ref{WW1}) and make use
 representation (\ref{tw-wf-gen})and equations
 of motion \p{tw-eq-1}, \p{tw-eq-2}, \p{tw-eq-3} written it terms
 of twistor variables
\begin{equation}
\label{equ-wf1}
i\,\pi_{\alpha}\,\frac{\partial}{\partial\rho_{\alpha}}\,\Psi^{(c)} =  \Psi^{(c)} \,,\qquad
i\,\bar\pi_{\dot\alpha}\,\frac{\partial}{\partial\bar\rho_{\dot\alpha}}\,\Psi^{(c)}
=  \Psi^{(c)} \,,
\end{equation}
\begin{equation}
\label{equ-wf2}
\left(\pi_{\alpha}\frac{\partial}{\partial \pi_{\alpha}} \ -\
\bar \pi_{\dot\alpha}\frac{\partial}{\partial \bar \pi_{\dot \alpha}} \ + \
\rho_{\alpha}\frac{\partial}{\partial \rho_{\alpha}} \ -\
\bar \rho_{\dot\alpha}\frac{\partial}{\partial \bar \rho_{\dot \alpha}}\right)\,
\Psi^{(c)} \ =\   2c \,  \Psi^{(c)}\,.
\end{equation}

\vspace{0.2cm}

Recall that the twistorial wave function {{$\Psi^{(c)}$}} is complex and therefore all component fields
$\psi^{(c\pm k)}(\pi,\bar \pi)$ in its expansion  are also complex.
In view of this
we must consider together with the field $\Psi^{(c)}$
 its complex conjugated field $(\Psi^{(c)})^*:=\bar\Psi^{(-c)}$
which has  the opposite charge $c \to -c$.

\section{\bf Twistor transform for infinite spin fields}

In this section, we establish a correspondence between twistor fields and fields defined in the four-dimensional Minkowski space-time.

For further convenience we introduce the dimensionless spinor
$$
\xi_{\alpha}:=M^{-1/2}\rho_{\alpha}\,,\qquad
\bar \xi_{\dot\alpha}:=M^{-1/2} \bar \rho_{\dot \alpha}\,.
$$
Then, the twistor wave function $\Psi^{(c)}$ of infinite integer-spin particle \p{tw-wf-gen} for $c=0$
can be represented in the form \cite{BFI}
\begin{eqnarray}
\Psi^{(0)}( \pi,\bar \pi;\xi,\bar\xi) &=&
\delta\left((\pi\xi)(\bar\xi\bar\pi)-M\right)\,
e^{\displaystyle -iq_0/p_0}\,
\hat\Psi^{(0)} (\pi,\bar \pi;\xi,\bar\xi)\,,
\\
&& \hat\Psi^{(0)} \ = \ \psi^{(0)}( \pi,\bar \pi) +
\sum\limits_{k=1}^{\infty} (\bar\xi\bar\pi)^{k}\,
\psi^{(k)}(\pi,\bar \pi) +
\sum\limits_{k=1}^{\infty} (\pi\xi)^{k}\,
\psi^{(-k)}(\pi,\bar \pi)\,. \nonumber
\end{eqnarray}
In the expansion of $\hat\Psi^{(0)}$,
 all components $\psi^{(k)}( \pi,\bar \pi)$ ($k\in \mathbb{Z}$)
in general are complex functions (fields).
Moreover, the quantity $p_0/q_0$ is expressed by means of the generalized Cartan-Penrose representations \p{tw-CP-1} in spinorial form as
$$
\frac{q_0}{p_0} \ = \
\frac{\sqrt{M}\sum\limits_{\alpha=\dot\alpha}(\pi_{\alpha}
\bar\xi_{\dot\alpha} + \xi_{\alpha}\bar\pi_{\dot\alpha})}
{\sum\limits_{\beta=\dot\beta}\pi_{\beta}\bar\pi_{\dot\beta}} \,.
$$

In the case {{$c=-1/2$}}, the wave function of the infinite half-integer
spin particle is
\begin{eqnarray}
\Psi^{(-\frac12)}( \pi,\bar \pi;\xi,\bar\xi) &=&
\delta\left((\pi\xi)(\bar\xi\bar\pi)-M\right)\,
e^{\displaystyle -iq_0/p_0}\,
\hat\Psi^{(-\frac12)} (\pi,\bar \pi;\xi,\bar\xi)\,,
\\
&& \hat\Psi^{(-\frac12)} \ = \ \psi^{(-\frac12)}( \pi,\bar \pi) +
\sum\limits_{k=1}^{\infty} (\bar\xi\bar\pi)^{k}\,
\psi^{(-\frac12+k)}(\pi,\bar \pi) +
\sum\limits_{k=1}^{\infty} (\pi\xi)^{k}\,
\psi^{(-\frac12-k)}(\pi,\bar \pi)\,. \nonumber
\end{eqnarray}
The expansion of the complex conjugated wave function
{{$\bar\Psi^{(+\frac12)}$}} has the form
\begin{eqnarray}
\bar\Psi^{(+\frac12)}( \pi,\bar \pi;\xi,\bar\xi) &=&
\delta\left((\pi\xi)(\bar\xi\bar\pi)-M\right)\,
e^{\displaystyle iq_0/p_0}\,
\hat{\bar\Psi}^{(+\frac12)} (\pi,\bar \pi;\xi,\bar\xi)\,,
\\
&& \hat{\bar\Psi}^{(+\frac12)} \ = \ \bar\psi^{(\frac12)}( \pi,\bar \pi) +
\sum\limits_{k=1}^{\infty} (\bar\xi\bar\pi)^{k}\,
\bar\psi^{(\frac12+k)}(\pi,\bar \pi) +
\sum\limits_{k=1}^{\infty} (\pi\xi)^{k}\,
\bar\psi^{(\frac12-k)}(\pi,\bar \pi) \, ,\nonumber
\end{eqnarray}
where the component fields $\bar\psi^{(r)}(\pi,\bar \pi)$ are complex conjugation of the
component fields {{$\psi^{(-r)}(\pi,\bar \pi)$}}:
$$
\left(\psi^{(-\frac12+k)}\right)^*=\bar\psi^{(\frac12-k)}\; ,
\qquad  k\in \mathbb{Z}\,.
$$

\subsection{The case of integer spins}

In this case the $\mathrm{U}(1)$-charge is zero, $c=0$,
and the space-time wave function is determined by means of the integral
Fourier transformation of twistor field $\Psi^{(0)} (\pi,\bar\pi;\xi,\bar\xi)$:
\begin{equation}\label{st-f-0}
\Phi(x;\xi,\bar\xi) \ = \ \int d^4 \pi \, e^{\displaystyle \,i p_{\alpha\dot\alpha} x^{\dot\alpha\alpha}}\,
\Psi^{(0)} (\pi,\bar\pi;\xi,\bar\xi) =
\int d^4 \pi \, e^{\displaystyle \,i \pi_{\alpha}\bar\pi_{\dot\alpha} x^{\dot\alpha\alpha}}\,
\Psi^{(0)} (\pi,\bar\pi;\xi,\bar\xi)
\,,
\end{equation}
where we have used the representation
$p_{\alpha\dot\alpha}= \pi_{\alpha}\bar\pi_{\dot\alpha}$ and perform integration
over the  measure $d^4 \pi :=
\frac12\, d\pi_{1}\wedge d\pi_{2}\wedge d\bar\pi_{\dot1}\wedge d\bar\pi_{\dot2}= d\phi \; d^4p \; \delta(p^2)$
(here $\phi$ is common phase in $\pi_\alpha$ which is not presented in $p_{\alpha\dot\alpha}=\pi_{\alpha}\bar\pi_{\dot\alpha}$).

\vspace{0.2cm}

\noindent
{\bf Proposition 4.} {\em The field $\Phi(x;\xi,\bar\xi)$ defined by the integral transformation \p{st-f-0} in coordinate
representation satisfies four equations}
\begin{equation}
 \label{eq00}
\begin{array}{c}
{\displaystyle \partial^{\alpha\dot\alpha}\partial_{\alpha\dot\alpha}\,
\Phi(x;\xi,\bar\xi) =  0 \,, \qquad
\left(i\frac{\partial}{\partial\xi_{\alpha}}
\partial_{\alpha\dot\alpha}
\frac{\partial}{\partial\bar\xi_{\dot\alpha}}-M\right)
\Phi(x;\xi,\bar\xi)  =  0 \,,}
\\ [0.2cm]
{\displaystyle \left(i\xi^{\alpha}\partial_{\alpha\dot\alpha}
\bar\xi^{\dot\alpha}+M\right)\Phi(x;\xi,\bar\xi)  =  0 \,,
\qquad
\left(\xi_{\alpha}\frac{\partial}{\partial\xi_{\alpha}}-
\bar\xi_{\dot\alpha}
\frac{\partial}{\partial\bar\xi_{\dot\alpha}}\right)
\Phi(x;\xi,\bar\xi)  =  0 \,.}
\end{array}
\end{equation}
{\bf Proof.} Make use the integral transformation \p{st-f-0}
 and equations of motion (\ref{equ-wf1}), (\ref{equ-wf2}) for $c=0$.

\subsection{The case of half-integer spins}

In this case the $\mathrm{U}(1)$-charge equals $c=- 1/2$.
Then we use the standard prescription of the twistorial
definition of space-time fields with nonvanishing helicities.
Namely, we have to insert the twistorial spinor
$\pi_\alpha$ in the integrand in the Fourier
transformation:
\begin{equation}
\label{s121}
\Phi_\alpha(x;\xi,\bar\xi)  =  \int d^4 \pi \, e^{\displaystyle \,i p_{\beta\dot\beta} x^{\dot\beta\beta}}\pi_\alpha\,
\Psi^{(-1/2)} (\pi,\bar\pi;\xi,\bar\xi) =
\int d^4 \pi \, e^{\displaystyle \,i \pi_{\beta}\bar\pi_{\dot\beta} x^{\dot\beta\beta}}\pi_\alpha\,
\Psi^{(-1/2)} (\pi,\bar\pi;\xi,\bar\xi)\,,
\end{equation}
and obtain the external spinor index {{$\alpha$}}.
Then the complex conjugate twistorial field with charge
$c=+1/2$ is defined analogously
\begin{equation}
\label{s122}
\bar\Phi_{\dot\alpha}(x;\xi,\bar\xi) \ = \ \int d^4 \pi \, e^{\displaystyle \,-i \pi_{\beta}\bar\pi_{\dot\beta} x^{\dot\beta\beta}}\,\bar\pi_{\dot\alpha}\,
\bar\Psi^{(+1/2)} (\pi,\bar\pi;\xi,\bar\xi)\,.
\end{equation}

{\bf Proposition 5.} {\em The space-time fields $\Phi_\alpha(x;\xi,\bar\xi)$ and $\bar\Phi_{\dot\alpha}(x;\xi,\bar\xi)$,
which correspond to the states with half-integer helicities, satisfy massless Dirac-Weyl equations
\begin{equation}
\label{eq01}
\partial^{\dot\alpha\alpha}\,\Phi_\alpha(x;\xi,\bar\xi)\ = \ 0 \, , \qquad
\partial^{\dot\alpha\alpha}\,\bar\Phi_{\dot\alpha}(x;\xi,\bar\xi) \ = \ 0 \, ,
\end{equation}
and integer spin equations:}
\begin{equation}
\label{eq02}
\begin{array}{ll}
{\displaystyle \left(i\xi^{\beta}\partial_{\beta\dot\beta}\bar\xi^{\dot\beta}
+M\right)
\Phi_\alpha(x;\xi,\bar\xi)  =  0 \,,} &
{\displaystyle \left(i\xi^{\beta}\partial_{\beta\dot\beta}\bar\xi^{\dot\beta}
-M\right)\bar\Phi_{\dot\alpha}(x;\xi,\bar\xi)  =  0 \,,}
\\ [7pt]
{\displaystyle \left(i\frac{\partial}{\partial\xi_{\beta}}\partial_{\beta\dot\beta}
\frac{\partial}{\partial\bar\xi_{\dot\beta}}-
M\right)\Phi_\alpha(x;\xi,\bar\xi)  =  0 \,, }&
{\displaystyle \left(i\frac{\partial}{\partial\xi_{\beta}}
\partial_{\beta\dot\beta}
\frac{\partial}{\partial\bar\xi_{\dot\beta}}
+M\right)\bar\Phi_{\dot\alpha}(x;\xi,\bar\xi) =  0 \, ,}
\\ [7pt]
{\displaystyle \left(\xi_{\beta}\frac{\partial}{\partial\xi_{\beta}}-
\bar\xi_{\dot\beta} \frac{\partial}{\partial\bar\xi_{\dot\beta}}\right)
\Phi_\alpha(x;\xi,\bar\xi) \ = \ 0 \, , } &
{\displaystyle \left(\xi_{\beta}\frac{\partial}{\partial\xi_{\beta}}-
\bar\xi_{\dot\beta}
\frac{\partial}{\partial\bar\xi_{\dot\beta}}\right)
\bar\Phi_{\dot\alpha}(x;\xi,\bar\xi) \ = \ 0 \,.}
\end{array}
\end{equation}
{\bf Proof.} Make use the integral transformations \p{s121}, \p{s122}
 and equations of motion (\ref{equ-wf1}), (\ref{equ-wf2}) for $c=\pm 1/2$.

 \vspace{0.2cm}

We stress that although the twistorial fields $\Psi^{(-1/2)} (\pi,\bar\pi;\xi,\bar\xi)$
and $\bar\Psi^{(+1/2)} (\pi,\bar\pi;\xi,\bar\xi)$ have nonvanishing charges
$c=\mp 1/2$, their integral transformed versions $\Phi_{\dot\alpha}(x;\xi,\bar\xi)$
and $\bar\Phi_{\dot\alpha}(x;\xi,\bar\xi)$ have zero $\mathrm{U}(1)$-charge.
This fact is crucial for forming infinite spin supermultiplets, as we will see below.

\section{\bf Infinite spin supermultiplet}

We unify fields $\Phi(x;\xi,\bar\xi)$ and $\Phi_\alpha(x;\xi,\bar\xi)$
with integer and half-integer helicities
into one supermultiplet. The fields $\Phi(x;\xi,\bar\xi)$ and $\Phi_\alpha(x;\xi,\bar\xi)$ contain
the bosonic $\psi^{(k)}(\pi,\bar \pi)$ and fermionic $\psi^{(k-1/2)}(\pi,\bar \pi)$
component fields ($k \in \mathbb{Z}$) with all
integer and half-integer spins, respectively.

It is natural to expect that
individual components $\psi^{(k)}(\pi,\bar \pi)$ and $\psi^{(k-1/2)}(\pi,\bar \pi)$ of these fields  should form the on-shell $\mathcal{N}{=}\,1$
higher spin supermultiplet. Therefore, the
bosonic (even) $\Phi(x;\xi,\bar\xi)$ and fermionic (odd)
$\Phi_\alpha(x;\xi,\bar\xi)$ fields themselves should form
the on-shell $\mathcal{N}{=}\,1$ infinite spin supermultiplet containing an infinite number of conventional supermultiplets.

Similar to the Wess-Zumino supermultiplet (see, e.g., \cite{WessB,BK}),
we write supersymmetry transformations of the fields $\Phi$ and $\Phi_\alpha$
in the form
\begin{equation}
\label{tr-susy}
\delta\,\Phi \ = \ \varepsilon^{\alpha}\Phi_\alpha \,, \qquad
\delta\,\Phi_\alpha \ = \ 2i\bar\varepsilon^{\dot\beta}\partial_{\alpha\dot\beta}\Phi \,,
\end{equation}
where $\varepsilon_{\alpha}$, $\bar\varepsilon_{\dot\alpha}$ are the constant odd Weyl spinors.
The commutators of these transformations are
\begin{equation}\label{com-susy}
\begin{array}{lcl}
\left(\delta_1\delta_2-\delta_2\delta_1\right)\Phi &=& -2i a^{\beta\dot\beta}\partial_{\beta\dot\beta}\Phi\,,
\\ [6pt]
\left(\delta_1\delta_2-\delta_2\delta_1\right)\Phi_\alpha &=& -2i a^{\beta\dot\beta}\partial_{\beta\dot\beta}\Phi_\alpha
+2ia_{\alpha\dot\beta}\partial^{\dot\beta\beta}\Phi_\beta\,,
\end{array}
\end{equation}
where
$a_{\alpha\dot\beta} \ := \ \varepsilon_{1\alpha}\bar\varepsilon_{2\dot\beta}- \varepsilon_{2\alpha}\bar\varepsilon_{1\dot\beta}$.
As we see, the superalgebra (\ref{com-susy}) is closed on-shell on the generator
$$
P_{\beta\dot\beta}=-i\partial_{\beta\dot\beta}
$$
due to the Dirac-Weyl equations of motion (\ref{eq01}).
Moreover, the whole system of equations of motion (\ref{eq00}), (\ref{eq01}),
(\ref{eq02}) is invariant
with respect to supersymmetry transformations  (\ref{com-susy}).

Using the inverse integral Fourier transformations,
we rewrite (\ref{com-susy})
as supersymmetry transformations for the twistor fields
$\Psi^{(0)} (\pi,\bar\pi;\xi,\bar\xi)$, $\Psi^{(-1/2)} (\pi,\bar\pi;\xi,\bar\xi)$ in the momentum representation:
\begin{equation}
\label{tr-susy-tw}
\delta\,\Psi^{(0)} \ = \ \varepsilon^{\alpha}\pi_\alpha \Psi^{(-1/2)}\,, \qquad
\delta\,\Psi^{(-1/2)} \ = \ -2\,\bar\varepsilon^{\dot\alpha}\pi_{\dot\alpha}\Psi^{(0)} \,.
\end{equation}
For the bosonic $\psi^{(k)}( \pi,\bar \pi)$
and fermionic $\psi^{(-\frac12+k)}( \pi,\bar \pi)$ twistorial components
at all $k\,{\in}\,\mathbb{Z}$ we have
\begin{equation}
\label{tr-susy-tw-comp}
\delta\,\psi^{(k)} \ = \ \varepsilon^{\alpha}\pi_\alpha \,\psi^{(-\frac12+k)}\,, \qquad
\delta\,\psi^{(-\frac12+k)} \ = \ -2\,\bar\varepsilon^{\dot\alpha}\pi_{\dot\alpha}\,\psi^{(k)} \,.
\end{equation}
The bosonic field $\psi^{(k)}$ and fermionic field $\psi^{(-\frac12+k)}$ at fixed
$k\,{\in}\,\mathbb{Z}$ describe massless states with helicities $(-k)$ and $(\frac12-k)$, respectively.
Thus, the infinite-component supermultiplet of the infinite spin stratifies into
an infinite number of levels with pairs of the fields
$\psi^{(k)}$, $\psi^{(-\frac12+k)}$ at fixed $k\,{\in}\,\mathbb{Z}$.
The supersymmetry transforms the bosonic and fermionic
fields into each other inside a given level $k$.
The boosts of the Poincare group transform the levels with
different $k$ and therefore mix the fields with different
values of $k$.

In final we point out that superfield description of infinite spin supermultiplet was presented in recent paper \cite{BuchGK}.

\vspace{1cm}

\section{Summary and outlook}

Let us summarize the obtained results.

\begin{itemize}

\item We have presented the new twistorial formulation of the massless infinite spin particles and fields.

\item We gave the helicity decomposition of twistorial infinite spin fields
and constructed the field twistor transform to define
the space-time infinite (continuous) spin fields
$\Phi(x;\xi,\bar\xi)$ and $\Phi_\alpha(x;\xi,\bar\xi)$.

\item We found the equations of motion for
$\Phi(x;\xi,\bar\xi)$ and $\Phi_\alpha(x;\xi,\bar\xi)$
and showed that these fields form the
$\mathcal{N}{=}\,1$ infinite spin supermultiplet.

\item A natural
question arises about status of such fields in Lagrangian field
theory and also about possibility to construct self-consistent
interaction of such fields. One of the commonly used methods for
this purpose is the BRST approach, which was used in the
case of continuous spin particles in \cite{Bengtsson13},
\cite{AlkGr} -
 \cite{Metsaev18a}.
In a recent paper \cite{BuchKrTak} the covariant Lagrangian
formulation of the infinite integer-spin field was constructed by
using the methods developed in \cite{BuchKrP,BuchKr}.

\end{itemize}

\section*{Acknowledgments}
S.F. and A.P.I. thank M.\,Podoinitsyn  for helpful discussion of results of this paper.
I.L.B. and A.P.I. acknowledge the partial support of the
Ministry of Science and High Education of Russian Federation,
project No.\,3.1386.2017. I.L.B. acknowledges the support of the Russian
Foundation for Basic Research, project No.\,18-02-00153.
A.P.I. acknowledges the support of the Russian Science Foundation, grant No.\,19-11-00131.


\begin{thebibliography}{96}

\bibitem{Slavnov}
A.A.\,Slavnov, {\it Ward identities in gauge theories}, Theor. Math.
Phys. {\bf 10}, 2 (1972) 99 (Teor. Mat. Fiz. {\bf 10}, 2 (1972)
153); {\it Invariant regularization of gauge theories},
Theor. Math. Phys., {\bf 13}, 2 (1972) 1064
(Teor. Mat. Fiz. {\bf 13}, 2 (1972) 174).

\bibitem{Wigner39}
E.P.\,Wigner,
{\it On unitary representations of the inhomogeneous Lorentz group},
Annals Math.  {\bf 40} (1939) 149.

\bibitem{Wigner47}
E.P.\,Wigner,
{\it Relativistische Wellengleichungen},
Z. Physik  {\bf 124} (1947) 665.

\bibitem{BargWigner}
V.\,Bargmann, E.P.\,Wigner,
{\it Group theoretical discussion of relativistic wave equations},
Proc. Nat. Acad. Sci. US  {\bf 34} (1948) 211.

\bibitem{Iv-Mack}
G.J.\,Iverson, G.\,Mack,
{\it Quantum fields and interactions of massless particles - the continuous spin case},
Annals Phys.  {\bf 64} (1971) 253.


\bibitem{BKRX}
L.\,Brink, A.M.\,Khan, P.\,Ramond, X.-Z. Xiong,
{\it Continuous spin representations of the Poincar\'{e} and superPoincar\'{e} groups},
J. Math. Phys.  {\bf 43} (2002) 6279, {\tt arXiv:hep-th/0205145}.

\bibitem{BekBoul}
X.\,Bekaert, N.\,Boulanger,
{\it The unitary representations of the Poincar\'{e} group in any spacetime dimension},
Lectures presented at
2nd Modave Summer School in Theoretical Physics,
6-12 Aug 2006, Modave, Belgium, {\tt arXiv:hep-th/0611263}.



\bibitem{BekMou}
X.\,Bekaert, J.\,Mourad,
{\it The continuous spin limit of higher spin field equations},
JHEP  {\bf 0601} (2006) 115, {\tt arXiv:hep-th/0509092}.

\bibitem{SchToro13a}
P.\,Schuster, N.\,Toro,
{\it On the theory of continuous-spin particles: wavefunctions and soft-factor scattering amplitudes},
JHEP  {\bf 1309} (2013) 104, {\tt arXiv:1302.1198\,[hep-th]}.

\bibitem{SchToro13b}
P.\,Schuster, N.\,Toro,
{\it On the theory of continuous-spin particles: helicity correspondence in radiation and forces},
JHEP  {\bf 1309} (2013) 105, {\tt arXiv:1302.1577\,[hep-th]}.

\bibitem{SchToro13c}
P.\,Schuster, N.\,Toro,
{\it A gauge field theory of continuous-spin particles},
JHEP  {\bf 1310} (2013) 061, {\tt arXiv:1302.3225\,[hep-th]}.

\bibitem{Bengtsson13}
A.K.H.\,Bengtsson,
{\it BRST Theory for Continuous Spin},
JHEP {\bf 1310} (2013) 108,
{\tt arXiv:1303.3799\,[hep-th]}.

\bibitem{SchToro15}
P.\,Schuster, N.\,Toro,
{\it A CSP field theory with helicity correspondence},
Phys. Rev.  {\bf D91} (2015) 025023, {\tt arXiv:1404.0675\,[hep-th]}.


\bibitem{Riv}
V.O.\,Rivelles,
{\it Gauge theory formulations for continuous and higher spin fields},
Phys. Rev.  {\bf D91} (2015) 125035, {\tt arXiv:1408.3576\,[hep-th]}.

\bibitem{BekNajSe}
X.\,Bekaert, M.\,Najafizadeh, M.R.\,Setare,
{\it A gauge field theory of fermionic Continuous-Spin Particles},
Phys. Lett. {\bf B760} (2016) 320, {\tt arXiv:1506.00973\,[hep-th]}.

\bibitem{Mets16}
R.R.\,Metsaev,
{\it Continuous spin gauge field in (A)dS space},
Phys. Lett.    {\bf B767} (2017) 458, {\tt arXiv:1610.00657\,[hep-th]}.

\bibitem{Mets17}
R.R.\,Metsaev,
{\it Fermionic continuous spin gauge field in (A)dS space},
Phys. Lett.    {\bf B773} (2017) 135, {\tt arXiv:1703.05780\,[hep-th]}.


\bibitem{Zin}
Yu.M.\,Zinoviev,
{\it Infinite spin fields in $d{=}\,3$ and beyond},
Universe {\bf 3} (2017) 63, {\tt  arXiv:1707.08832\,[hep-th]}.

\bibitem{BekSk}
X.\,Bekaert, E.D.\,Skvortsov,
{\it Elementary particles with continuous spin},
Int. J. Mod. Phys.   {\bf A32} (2017) 1730019, {\tt arXiv:1708.01030\,[hep-th]}.

\bibitem{HabZin}
M.V.\,Khabarov, Yu.M.\,Zinoviev,
{\it Infinite (continuous) spin fields in the frame-like formalism},
Nucl. Phys. {\bf B928} (2018) 182, {\tt arXiv:1711.08223\,[hep-th]}.

\bibitem{AlkGr}
K.B.\,Alkalaev, M.A.\,Grigoriev,
{\it Continuous spin fields of mixed-symmetry type},
JHEP {\bf 1803} (2018) 030, {\tt arXiv:1712.02317\,[hep-th]}.


\bibitem{Metsaev18}
R.R.\,Metsaev,
{\it BRST-BV approach to continuous-spin field},
Phys. Lett. {\bf B781} (2018) 568,
{\tt arXiv:1803.08421\,[hep-th]}.


\bibitem{BuchKrTak}
I.L.\,Buchbinder, V.A.\,Krykhtin, H.\,Takata,
{\it BRST approach to Lagrangian construction for bosonic continuous spin field},
Phys. Lett.   {\bf B785} (2018) 315,
{\tt arXiv:1806.01640\,[hep-th]}.

\bibitem{ACG18}
K.\,Alkalaev, A.\,Chekmenev, M.\,Grigoriev,
{\it Unified formulation for helicity and continuous spin fermionic fields},
JHEP {\bf 1811} (2018) 050,
{\tt arXiv:1808.09385\,[hep-th]}.

\bibitem{Metsaev18a}
R.R.\,Metsaev,
{\it Cubic interaction vertices for massive/massless continuous-spin fields and arbitrary spin fields},
JHEP {\bf 1812} (2018) 055,
{\tt arXiv:1809.09075\,[hep-th]}.



\bibitem{Riv18}
V.O.\,Rivelles,
{\it A gauge field theory for continuous spin tachyons},
{\tt arXiv:1807.01812\,[hep-th]}.

\bibitem{Metsaev19}
R.R.\,Metsaev, {\it Light-cone continuous-spin field in AdS space},
 Phys. Lett. {\bf B 793} (2019) 134;
 {\tt arXiv:1903.10495\,[hep-th]}.

\bibitem{Vas1989}
M.A.\,Vasiliev,
{\it Consistent equations for interacting massless fields of all spins in the first order in curvatures},
Annals Phys. {\bf 190} (1989) 59.

\bibitem{Vas1991}
M.A.\,Vasiliev,
{\it Algebraic  aspects  of  the  higher  spin  problem},
Phys. Lett. {\bf B257} (1991) 111.


\bibitem{Vas1992}
M.A.\,Vasiliev,
{\it More on equations of motion for interacting massless fields of all spins in (3+1)-dimensions},
Phys. Lett. {\bf B285} (1992) 225.


\bibitem{Vas2001}
M.A.\,Vasiliev,
{\it Progress in higher spin gauge theories},
{Proceedings of the International Conference on Quantization, Gauge Theory, and Strings:
Conference Dedicated to the Memory of Prof. E.\,Fradkin,}
Eds. A.\,Semikhatov, M.\,Vasiliev, V.\,Zaikin, Scientific World, Moscow, 2001, 452-472, {\tt arXiv: hep-th/0104246}.


\bibitem{Vas2002}
M.A.\,Vasiliev,
{\it Relativity, causality, locality, quantization and duality in the $Sp(2M)$ invariant generalized spacetime},
in {Multiple Facets of Quantization and Supersymmetry}, Michael Marinov Memorial Volume,
Eds. M.\,Olshanetsky and A.\,Vainshtein, World Scientific, 2002,
826-872,
{\tt arXiv:hep-th/0111119}.

\bibitem{Vas2005}
X.\,Bekaert, S.\,Cnockaert, C.\,Iazeolla, M.A.\,Vasiliev,
{\it Nonlinear higher spin theories in various dimensions},
{Proceedings of the 1st Solvay Workshop on Higher Spin Gauge Theories, 12-14 May 2004. Brussels, Belgium},
Eds. R.\,Argurio, G.\,Barnich, G.\,Bonelli, M.\,Grigoriev, Int. Solvay Institutes, 2006,
132-197,
{\tt arXiv:hep-th/0503128}.

\bibitem{MundSY}
J.\,Mund, B.\,Schroer, J.\,Yngvason,
{\it String localized quantum fields from Wigner representations},
Phys. Lett.  {\bf B596} (2004) 156, {\tt arXiv:math-ph/0402043}.


\bibitem{Vas2018}
M.A.\,Vasiliev,
{\it From Coxeter higher-spin theories to strings and tensor models},
{\tt arXiv:1804.06520\,[hep-th]}.




\bibitem{BFIR}
I.L.\,Buchbinder, S.\,Fedoruk, A.P.\,Isaev, A.\,Rusnak,
{\it Model of massless relativistic particle with continuous spin and its twistorial description},
JHEP  {\bf 1807} (2018) 031,
{\tt arXiv:1805.09706\,[hep-th]}.

\bibitem{BFI}
I.L.\,Buchbinder, S.\,Fedoruk, A.P.\,Isaev,
{\it Twistorial and space-time descriptions of massless
infinite spin (super)particles and fields},
Nucl. Phys. B {\bf 945} (2019) 114660, {\tt arXive:1903.07947[hep-th]}.





\bibitem{Pen67}
R.\,Penrose,
{\it Twistor algebra},
J. Math. Phys. {\bf 8} (1967) 345.

\bibitem{PC72}
R.\,Penrose, M.A.H.\,MacCallum,
{\it Twistor theory: an approach to the quantization of fields and spacetime},
Phys. Rept. {\bf 6} (1972) 241.


\bibitem{PenRin}
R.\,Penrose, W.\,Rindler,
{\it Spinors And Space-time. Vol. 2: Spinor And Twistor Methods In Space-time Geometry},
Cambridge University Press, 1988, 512 pages.


\bibitem{WessB}
J.\,Wess, J.\,Bagger,
{\it Supersymmetry and supergravity},
Princeton, NJ, USA: Princeton University Press, 1992, 259 pages.


\bibitem{BK}
I.L.\,Buchbider, S.M.\,Kuzenko, {\it Ideas and Methods of Supersymmetry
and Supergravity}, IOP Publ., 1998, 656 pages.

\bibitem{BuchGK}
I.L.\,Buchbinder, S.J.\,Gates, K.\,Koutrolikos,
{\it Superfield continuous spin equations of motion},
Phys. Lett. {\bf B793} (2019) 445, {\tt arXiv:1903.08631\,[hep-th]}.

\bibitem{BuchKrP}
I.L.\,Buchbinder, V.A.\,Krykhtin, A.\,Pashnev,
{\it BRST approach to Lagrangian construction for fermionic massless higher spin fields},
Nucl. Phys. {\bf B711} (2005) 367, {\tt arXiv:hep-th/0410215}.

\bibitem{BuchKr}
I.L.\,Buchbinder, V.A.\,Krykhtin,
{\it Gauge invariant Lagrangian construction for massive bosonic higher spin fields in D dimensions},
Nucl. Phys. {\bf B727} (2005) 537, {\tt arXiv:hep-th/0505092}.






\end{thebibliography}
\end{document}